\documentclass[aip,jcp,11pt,reprint,floatfix,superscriptaddress]{revtex4-1}
\usepackage[utf8]{inputenc}
\usepackage{amsmath}
\usepackage{color}
\usepackage{graphicx}

\DeclareMathOperator{\Tr}{Tr} 

\begin{document}
\title{A coupled cluster framework for electrons and phonons}
\date{March 2019}
\author{Alec F. White}
\thanks{These two authors contributed equally to this work}
\affiliation{Division of Chemistry and Chemical Engineering, California Institute of Technology, Pasadena,
California 91125, USA}
\email{whiteaf@berkeley.edu}

\author{Yang Gao}
\thanks{These two authors contributed equally to this work}
\affiliation{Division of Engineering and Applied Science, California Institute of Technology, Pasadena,
California 91125, USA}

\author{Austin J. Minnich}
\affiliation{Division of Engineering and Applied Science, California Institute of Technology, Pasadena,
California 91125, USA}

\author{Garnet Kin-Lic Chan}
\email{gkc1000@gmail.com}
\affiliation{Division of Chemistry and Chemical Engineering, California Institute of Technology, Pasadena,
California 91125, USA}

\begin{abstract}
We describe a coupled cluster framework for coupled systems of electrons and phonons. Neutral and charged excitations are accessed via the equation-of-motion version of the theory. Benchmarks on the Hubbard-Holstein model allow us to assess the strengths and weaknesses of different coupled cluster approximations which generally perform well for weak to moderate coupling. Finally, we report progress towards an implementation for {\it ab initio} calculations on solids, and present some preliminary results on finite-size models of diamond. We also report the implementation of electron-phonon coupling matrix elements from crystalline Gaussian type orbitals (cGTO) within the PySCF program package.
\end{abstract}
\maketitle

\section{Introduction}\label{sec:intro}
The electron-phonon interaction (EPI) underlies a vast array of phenomena in materials science and condensed matter physics. For example, the temperature dependence of electronic transport and optical properties can be largely traced to these interactions between electronic and nuclear degrees of freedom, while EPIs are the key interaction underpinning the Bardeen-Cooper-Schrieffer theory of  superconductivity.

The phenomenology surrounding EPIs has been extensively studied in the context of various lattice models and semi-empirical Hamiltonians. For example, the Hamiltonians of Fr\"{o}lich\cite{Frohlich1954} and Holstein\cite{Holstein1959} capture the limits of non-local and local electron-phonon interactions respectively. The Su, Schreiffer, and Heeger (SSH) model was introduced as a simplified model of 1-dimensional polyacetylene including EPIs\cite{Su1979}, and is now commonly used as a simple example of a 1-dimensional system with topological character\cite{Hasan2010}. An electron-interaction term, usually in the form of a Hubbard interaction, is often added to the above models to study the regime where both EPI and electron-electron repulsion are important. The Hubbard-Holstein (HH) model is perhaps the simplest such model and its well-studied phase diagram clearly displays the rich structure that can result from the subtle interplay of electron-phonon and electron-electron interactions\cite{Beni1974,Berger1995,Bauer2010,Nowadnick2012}.

Complementary to the study of model Hamiltonians has been the development of an {\it ab initio} theory of EPIs. This theory is reviewed in Refs.~\onlinecite{Giustino2017,Bernardi2016}. Typically, one uses density functional theory (DFT)\cite{Parr1989,Martin2004} for the electronic structure and then computes the EPIs using finite difference differentiation (the ``supercell approach")\cite{Dacorogna1985,Dacorogna1985a,Lam1986} or density functional perturbation theory (DFPT)\cite{Baroni1987,Gonze1992,Savrasov1992}. While the expense of these calculations often necessitates the use of DFT, there have been some attempts to move beyond the DFT framework\cite{Lazzeri2008,Faber2011,Yin2013,Antonius2014,Faber2015,Monserrat2016,Li2019}. These results suggest that going beyond DFT quasiparticle energies can change the effects of the EPI significantly. Converging any calculation with respect to the grid used for integration over the Brillouin zone has required the development of specialized interpolation schemes so that the EPI matrix elements may be represented on a very dense grid\cite{Giustino2007}. Because of the large size of this quantity, calculations of observables have been limited to relatively simple expressions. This is in contrast to the situation for model problems where, in most regimes, the coupled problem is solved nearly exactly for a small number of degrees of freedom.

Our goal is to eventually bridge the gap between the sophisticated treatment of simplified EPIs typical in model problems and simple treatments using {\it ab initio} EPIs. Our tool will be coupled cluster (CC) theory which has long formed the basis of some of the most accurate calculations of molecular electronic structure\cite{Coester1958,Cizek1966,Paldus1975,Cizek1980,Bartlett2007,Shavitt2009}. CC theory has also been extended to treat the electronic structure of periodic systems\cite{Hirata2001,Hirata2004,Gruneis2011,McClain2017,Motta2017,Gruber2018} and the vibrational structure of molecules including anharmonicity\cite{Prasad1988,Nagalakshmi1994,MadhaviSastry1994,Latha1996,DurgaPrasad2002,Banik2008,Banik2010,Christiansen2004,Seidler2007,Seidler2009}. In the spirit of this work, Monkhorst has suggested a ``molecular coupled cluster" method\cite{Monkhorst1987} which seeks to use CC theory for coupled electrons and nuclei in molecules when the Born-Oppenheimer approximation breaks down. The equation-of-motion (EOM) formalism is one way to obtain excited state properties from a coupled cluster ground state\cite{Stanton1993, Nooijen1995, Krylov2008}. Though most commonly used for molecular electronic excited states, EOM-CC methods have been applied to excitations in periodic solids\cite{McClain2016,McClain2017,Gao2020,Wang2020} as well as to vibrational excited states in molecules\cite{Nagalakshmi1994,Seidler2007,Banik2008,Faucheaux2015}.  In this work we describe a coupled cluster theory and corresponding EOM extension for interacting electrons and phonons. This theory is similar to some coupled cluster theories for cavity polaritons that have been independently developed over the last year\cite{Mordovina2020,Haugland2020}.

In Section~\ref{sec:theory} we present a coupled cluster theory for electrons and phonons that treats the interacting problem at a correlated level of theory. We discuss the derivation and implementation of the equations for different ground-state and excited-state methods. In Section~\ref{sec:bench} we apply the method to the Hubbard-Holstein model. We find that CC methods generally perform well for weak to moderate electron-phonon coupling but break down for strong coupling using standard (fixed-particle-number) reference states. Finally we describe work towards an {\it ab initio} implementation and present some initial calculations on the zero-point renormalization (ZPR) of the band gap of diamond. These calculations allow us to identify some of the difficulties in applying coupled cluster theory to the {\it ab initio} problem. In particular, the large finite-size error suggests that sampling the Brillouin zone more completely, for example by more approximate, perturbative calculations, is necessary for quantitative accuracy. However, the coupled cluster framework presented here provides the means to evaluate such approximations and relax them when necessary.

\section{Theory}\label{sec:theory}
In what follows we will use $a^{\dagger}$ ($a$) to represent fermionic creation (annihilation) operators and $b^{\dagger}$ ($b$) to represent bosonic creation (annihilation) operators. Though we focus on the case of electrons and phonons, the formalism can be applied to any system of interacting fermions and bosons. 

\subsection{Coupled cluster theory for fermions and bosons}
The coupled cluster method for fermions can be derived from an exponential wavefunction ansatz
\begin{equation}
	|\Psi_{CC}\rangle = e^T|\Phi_0\rangle,
\end{equation}
where $|\Phi_0\rangle$ is a single determinant reference. 
The $T$-operator is defined in some space of excited configurations such that
\begin{equation}
	T = \sum_{ia} t_{i}^a a_a^{\dagger}a_{i} + 
	\frac{1}{4}\sum_{ijab}t_{ij}^{ab} a_a^{\dagger}a_b^{\dagger}a_ja_i + \ldots 
\end{equation}
where $i$ and $a$ index occupied and virtual orbitals respectively.

Generally, the $T$-operator is truncated at some finite excitation level. For example, letting $T = T_1 + T_2$ yields the coupled cluster singles and doubles (CCSD) approximation. The coupled cluster energy and amplitudes are then determined from a projected Schrodinger equation:
\begin{align}
	\langle \Phi_0|e^{-T}He^T|\Phi_0\rangle &= E_{\mathrm{HF}} + E_{\mathrm{CC}} \\
	\langle \Phi_{\mu}|e^{-T}He^T|\Phi_0\rangle &= 0.
\end{align}

Two different flavors of bosonic coupled cluster have been used in the past for vibrational excitations:
\begin{enumerate}
    \item excitations in each mode are treated as bosons such that the $n$th excited state is an occupation of $n$ bosons\cite{Prasad1988,Faucheaux2015}
    \item each excited state in each mode is treated as a separate bosonic degree of freedom with the constraint that exactly one state in each mode is occupied\cite{Christiansen2004}.
\end{enumerate}

The differences between these two pictures have been discussed in Ref.~\onlinecite{Christiansen2004a}. When formulating coupled cluster theory, (1) has the advantage that no truncation of the excitation space beyond the truncation of the $T$ operator is necessary. This means that $e^T$ acting on the vacuum creates up to infinite order excitations while it is only parameterized by a finite number of operators. On the other hand, (2) has the advantage that more general ``modals" can be used, or, to put in another way, the reference need not be harmonic. Both formulations have been used in vibrational coupled cluster theories\cite{Prasad1988,Christiansen2004,Faucheaux2015}, and both pictures have been used recently in independent works on coupled cluster methods for molecules interacting with cavity photons\cite{Haugland2020,Mordovina2020}. Since we will be confining ourselves to the harmonic approximation anyway, we will use  second quantization of type (1):
\begin{equation}
    |\Psi_{\text{CC}}\rangle = e^T|0\rangle
\end{equation}
\begin{equation}
    T = \sum_{x} t_{x} b_{x}^{\dagger} + 
    \frac{1}{2}\sum_{xy}t_{xy}b_{x}^{\dagger}b_{y}^{\dagger} + \ldots
\end{equation}
where we have used $x,y,\ldots$ to index the bosonic modes.

To construct a coupled cluster formalism for electron-phonon systems, we use an exponential ansatz on top of a product reference:
\begin{equation}
    |\Psi_{\mathrm{CC}}\rangle  = e^T|\Phi_0\rangle |0\rangle .
\end{equation}
We will refer to theories of this type as electron-phonon coupled cluster (ep-CC).

\subsection{Coupled cluster models for electron-phonon systems}
In general, the $T$ operator for the coupled theory consists of a purely electronic part, purely phononic part, and a coupled part:
\begin{equation}
    T = T_{\mathrm{el}} + T_{\mathrm{ph}} + T_{\mathrm{ep}}.
\end{equation}
The level at which we truncate each of these pieces determines the accuracy of the method. We will use SDT$\ldots$ to specify the electronic amplitudes as is common for electronic coupled cluster, and we will use numbers, 123$\ldots$, to indicate the purely phononic amplitudes that we include. A combination of letters and numbers are used for the coupled amplitudes. The theories considered in this work are shown in Table~\ref{tab:models}.
\begin{table*}
    \centering
    \begin{tabular}{c|cc}
    \hline\hline
        model &  $T_{\mathrm{ph}}$ &$T_{\mathrm{ep}}$\\ \hline
        ep-CCSD-1-S1 &$ $ $\sum_{x} t_{x} b_{x}^{\dagger}$ & 
        $\sum_{ia,x} t_{i,x}^ab_x^{\dagger}a_a^{\dagger}a_i$\\
        ep-CCSD-12-S1 & $\sum_{x} t_{x} b_{x}^{\dagger} + 
    \frac{1}{2}\sum_{xy}t_{xy}b_{x}^{\dagger}b_{y}^{\dagger}$& 
    $\sum_{ia,x} t_{i,x}^ab_x^{\dagger}a_a^{\dagger}a_i$\\
        ep-CCSD-12-S12 &$\sum_{x} t_{x} b_{x}^{\dagger} + 
    \frac{1}{2}\sum_{xy}t_{xy}b_{x}^{\dagger}b_{y}^{\dagger}$  & 
    \hspace{22pt}$\sum_{ia,x} t_{i,x}^ab_x^{\dagger}a_a^{\dagger}a_i + \frac{1}{2}\sum_{ia,xy}t_{i,xy}^ab_x^{\dagger}b_y^{\dagger}a_a^{\dagger}a_i$\\ 
    \hline\hline
    \end{tabular}
    \caption{The names, phonon, and electron-phonon amplitudes for the theories considered in this work. All the theories include singles and doubles for the pure electronic part of the amplitudes (not shown).}
    \label{tab:models}
\end{table*}
The theories considered here all have a computational scaling of $N^6$ where $N$ is the system size assuming that the numbers of occupied orbitals, virtual orbitals, and phonon modes all scale with the system size, $N$. Note that our ep-CCSD-1-S1 method is the same as the QED-CCSD-1 method presented in Ref.~\onlinecite{Haugland2020}.

For the simplest theory, ep-CCSD-1-S1, the energy and amplitude equations were derived by hand diagrammatically as described in Appendix~\ref{app:diagram}. For the more complicated theories, we used a code generator which is described in Appendix~\ref{app:codegen}.

\subsection{Equation of motion coupled cluster for excitations}
Excited states can be computed within the EOM formalism which parameterizes a neutral or charged excitation by applying an excitation operator to the CC ground state:
\begin{equation}
    |R\rangle = R|\Psi_{\mathrm{CC}}\rangle = 
    Re^T|\Phi_0, 0\rangle.
\end{equation}
Because the excitation operator, $R$, commutes with the excitation operators in $T$, solving this eigenvalue problem is equivalent to finding a right eigenvector of the similarity transformed Hamiltonian:
\begin{equation}
    \langle \mu|\bar{H}R^n|\Phi_0, 0\rangle = E_nR_{\mu}^n.
\end{equation}
Here, $E_n$ is the energy of the $n$th excited state, and $\mu$ indexes an element of the excitation operator $R$.

The excitation operator, $R$, can be chosen to access charged or neutral excitations:
\begin{widetext}
\begin{align}
    R_{\mathrm{EE}} &= \sum_{ia}r_i^aa_a^{\dagger}a_i 
    + \frac{1}{4}\sum_{ijab} r_{ij}^{ab}a_a^{\dagger}a_b^{\dagger}a_ja_i
    + \sum_xr_xb_x^{\dagger} + \sum_{ia,x}r_{i,x}^ab_x^{\dagger}
    a_a^{\dagger}a_i\\
    R_{\mathrm{IP}} &= \sum_i r_ia_i + \frac{1}{2}\sum_{ija}
    r_{ij}^aa_a^{\dagger}a_ia_j + \sum_{ix}r_{ix}b_x^{\dagger}a_i\\
    R_{\mathrm{EA}} &= \sum_a r^a a_a^{\dagger} 
    + \frac{1}{2}\sum_{iab}r_{i}^{ab}a_b^{\dagger}a^{\dagger}_aa_i
    + \sum_{ax}r_x^ab^{\dagger}_x a_a^{\dagger}
\end{align}
\end{widetext}

In practice, the eigenvalue problem is solved by iterative diagonalization. The necessary equations for the sigma vectors are derived and implemented efficiently as described in Appendix~\ref{app:EOM}.

\section{Benchmarks: the Hubbard-Holstein model}\label{sec:bench}
In order to understand the strengths of this method, we will study the Hubbard-Holstein (HH) model, a simple lattice model of correlated electrons and phonons. The Hubbard-Holstein Hamiltonian is
\begin{widetext}
\begin{equation}
	H = -t\sum_{j\sigma}\left(a_{(j + 1)\sigma}^{\dagger}a_{j\sigma} + \mathrm{h.c.}\right) + U\sum_{j}n_{j\uparrow}n_{j\downarrow} 
	+ \omega \sum_Jb_Jb_J^{\dagger} + g\sum_{j}n_j(b_J + b_J^{\dagger}) 
\end{equation}
\end{widetext}
where the lowercase and capital indices run over the fermionic and bosonic degrees of freedom at each lattice site and $\sigma$ runs over the spin degrees of freedom of the fermions. The fermionic part of the Hamiltonian is a Hubbard model with hopping $t$ and on-site repulsion $U$. The bosonic part of the Hamiltonian is an independent oscillator at each site with frequency $\omega$, and the final term couples the fermionic density at a given site with a linear displacement in the oscillator at that site. This coupling is controlled by $g$.

The HH model is an important model in condensed matter physics as it captures both antiferromagnetic order due to electron correlation and pairing from the electron-phonon interaction\cite{Beni1974,Guinea1983,Hirsch1983,Caron1984,Hirsch1985,Zheng1990,Berger1995,Yonemitsu1996,LaMagna1997,Pao1998,Clay2005,Koller2005,Hardikar2007,Tezuka2007,Bauer2010,Nowadnick2012,Hohenadler2013,Murakami2013,Costa2020}. As a minimal model of electron correlation and electron-phonon coupling, it is an ideal benchmark with which we can evaluate the performance of our coupled cluster models in different regimes. The electron-phonon coupling strength,
\begin{equation}
    \lambda \equiv \frac{g^2}{\omega},
\end{equation}
provides a measure of the effective strength of the electron-phonon interaction. Using a path-integral framework, the phonon degrees of freedom can be integrated out to yield an effective electron-electron interaction, the static limit of which becomes attractive when \begin{equation}
    \lambda = \frac{U}{2}.
\end{equation}
Note that our definition of $\lambda$ may differ by a factor of 2 from some other common definitions. For large coupling, the effective electron-electron interaction is attractive, and we would not expect our coupled cluster methods to perform well for such an attractive interaction. The extension to this regime should be possible by breaking particle number symmetry\cite{Duguet2017,Qiu2019}, but this is beyond the scope of this work.

\subsection{The four-site HH model: Benchmark of ground-state methods}
\begin{figure*}
    \centering
    \includegraphics[scale=0.9]{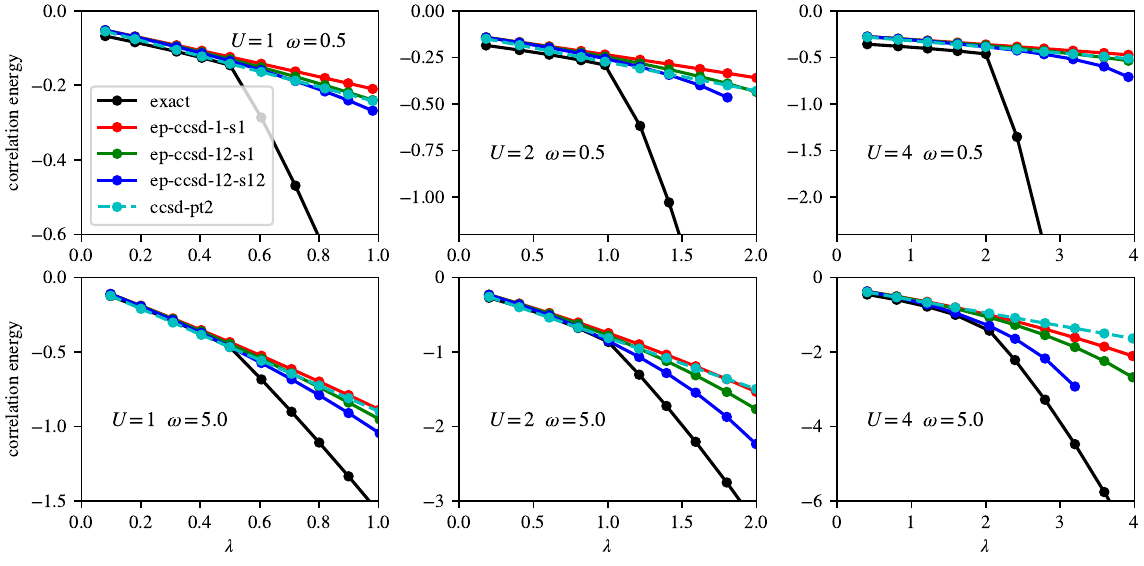}
    \caption{Correlation energy of the 4-site HH model for $U = 1,2,4$ and $\omega = 0.5, 5.0$. In all cases there is a qualitative change at $\lambda = 0.5U$ which is not captured by the approximate methods presented here. Both the energy and the coupling strength $\lambda$ are plotted in units of the hopping, $t$.}
    \label{fig:hh4}
\end{figure*}

The four-site (linear) HH model at half-filling is numerically solvable by exact diagonalization. In Figure~\ref{fig:hh4} we compare the correlation energy from three CC methods (see Table~\ref{tab:models}) with the exact correlation energy. In all cases, we use an unrestricted Hartree-Fock (UHF) reference and a generalized coherent state reference for the oscillators,
\begin{equation}
    \tilde{b}_I = b_I + g\frac{\langle \Phi_0|n_i|\Phi_0\rangle}{\omega},
\end{equation}
where $\Phi_0$ is the fermionic UHF reference. In terms of these transformed boson operators, the Hamiltonian has the same form except that the interaction term appears as
\begin{equation}
    g\sum_{j}(n_j - \langle n_j\rangle)(\tilde{b}_J + \tilde{b}_J^{\dagger})
\end{equation}
and there is an energy shift of
\begin{equation}
    -g^2\sum_i\frac{\langle n_i \rangle^2}{\omega}.
\end{equation}
This transformation diagonalizes the effective phononic Hamiltonian obtained by normal ordering the electronic part of the EPI term.

In addition to the coupled cluster methods, we also show the energy computed by adding a 2nd order perturbation theory (PT2) correction to the fermionic CCSD energy. The correction is given, in the UHF orbital basis, as
\begin{equation}\label{eqn:pt2}
    E_{\mathrm{pt2}} = -\sum_{ia,I} \frac{ |g_{i,I}^a|^2}
    {\varepsilon_a - \varepsilon_i + \omega}
\end{equation}
where $i$ ($a$) are occupied (virtual) orbitals and $I$ runs over the oscillators. Note that the interaction, $g$, becomes a generally non-diagonal tensor in the UHF orbital basis.

The correlation energy computed from these methods is compared to the exact results in Figure~\ref{fig:hh4} for $U = 1, 2, 4$ and $\omega = 0.5, 5.0$. The values of $U$ are chosen to be low enough that CCSD should provide qualitatively correct results in the limit of zero EPI, while the two values of $\omega$ are chosen to show approximately the limits of low frequency (adiabatic) and high frequency (anti-adiabatic). The transition to an attractive effective potential at $\lambda = U/2$ is evident in all cases, and the approximate methods described here fail qualitatively above this transition as expected.

For $\lambda < U/2$, all the methods shown here provide qualitatively correct results in the adiabatic and anti-adiabatic limits. The coupled cluster methods are systematic in that ep-CCSD-12-S12 outperforms ep-CCSD-12-S1 which outperforms ep-CCSD-1-S1 in all cases. This is one of the primary advantages of coupled cluster theory. The CCSD-PT2 method performs surprisingly well on this problem because Equation~\ref{eqn:pt2} tends to overestimate the electron-phonon correlation energy while CCSD underestimates the electronic correlation energy.

\subsection{EOM-ep-CCSD-1-S1 for charge and spin gaps}
In the thermodynamic limit, the 1-dimensional HH model at half-filling has a well-studied phase diagram: a Mott phase at small $\lambda/U$, a Peierls phase at large $\lambda/U$, and a metallic phase in between\cite{Clay2005,Koller2005,Tezuka2007,Hardikar2007,Werner2007,Bauer2010,Hohenadler2013}.

\begin{figure}
    \centering
    \includegraphics{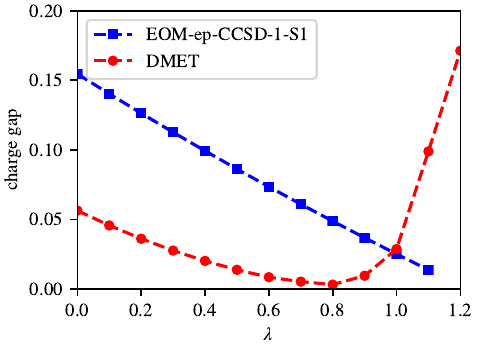}
    \caption{Charge gap of the HH model in the thermodynamic limit for $U = 1.6$ and $\omega = 0.5$. At $\lambda = 0.8$ ep-CCSD-1-S1 will break down and we would not expect correct results for $\lambda > 0.8$. The density matrix embedding theory (DMET) results are from Ref.~\onlinecite{Reinhard2019}.}
    \label{fig:hhc_ad}
\end{figure}
\begin{figure}
    \centering
    \includegraphics{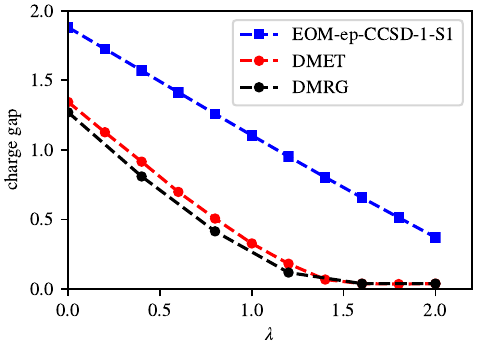}
    \caption{Charge gap of the HH model in the thermodynamic limit for $U = 4$ and $\omega = 5.0$. At $\lambda = 2.0$ ep-CCSD-1-S1 will break down and we would not expect correct results for $\lambda > 2.0$. The density matrix embedding theory (DMET) and density matrix renormalization group (DMRG) calculations are from Ref.~\onlinecite{Reinhard2019}.}
    \label{fig:hhc_anti}
\end{figure}

In Figures~\ref{fig:hhc_ad} and~\ref{fig:hhc_anti} we show the charge gap computed by IP/EA-EOM-ep-CCSD-1-S1 in the adiabatic case and the anti-adiabatic case respectively. In Figure~\ref{fig:hhc_ad} we show the extrapolated EOM-ep-CCSD-1-S1 band gap for $\omega = 0.5$ and $U = 1.6$. The extrapolation uses $L = 64$ and $L = 128$ systems with periodic boundary conditions and assumes asymptotically $1/L$ behavior. EOM coupled cluster often performs poorly on systems that are nearly metallic as can be seen at $\lambda = 0$ (the $U = 1.6$ Hubbard model). The results are qualitatively correct for $\lambda < 0.6$ though the closing of the gap at $\lambda = 0.6$ and Peierls insulator at $\lambda > 0.6$ are not captured by this approximation. In particular, note that EOM-ep-CCSD-1-S1 for the HH model does not perform worse than EOM-CCSD for the Hubbard model. In the anti-adiabatic case (Figure~\ref{fig:hhc_anti}: $\omega = 5.0$ and $U = 4$) the performance is similar. In this case, finite-size effects are less pronounced, and an extrapolation from calculations on $L = 32$ and $L = 64$ lattices is sufficient to estimate the thermodynamic limit. Because of the larger $U$, the model has a larger gap at $\lambda = 0$ and it is less severely overestimated by EOM. Again, EOM is qualitatively correct for small $\lambda$, but it breaks down as the system becomes metallic.

\begin{figure}
    \centering
    \includegraphics{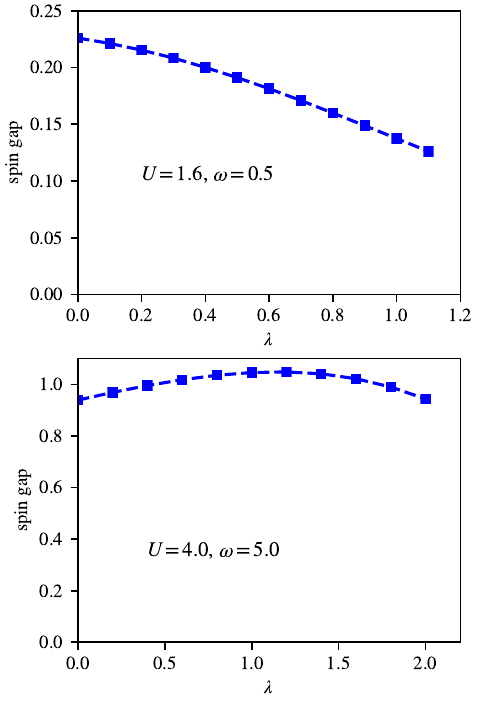}
    \caption{Spin gap of the Hubbard-Holstein model for $\omega = 0.5$ (top) and $\omega = 5.0$ (bottom) computed with EOM-CCSD-1-S1.}
    \label{fig:spin}
\end{figure}

In Figure~\ref{fig:spin} we show the spin gap as a function of $\lambda$ computed with EOM-EE-CCSD-1-S1.
 Though we expect EOM to overestimate the spin gap, it should be qualitatively correct for $\lambda < U/2$. Note that the spin gap is larger than the charge gap in both cases, and, unlike the charge gap, it does not appear to be going to zero. This is consistent with the non-zero spin gap observed in the intermediate metallic phase for this model\cite{Hohenadler2013} although it is not clear that coupled cluster theory provides the proper description of the underlying intermediate phase. 
 
\section{Application to \textit{ab initio} calculations of periodic solids}
The extension to {\it ab initio} problems requires a  Hamiltonian of the form
\begin{equation}
    H = H_{\mathrm{el}} + H_{\mathrm{ph}} + H_{\mathrm{ep}}
\end{equation}
where $H_{\mathrm{ep}}$ is both detailed enough to capture the physics of electron-phonon coupling from first principles and simple enough so that the matrix elements can be easily computed in the relevant basis. As we describe in Section~\ref{sec:ai_epc}, this is already quite a challenge. This is further complicated by the expense of controlling finite-size errors. In Section~\ref{sec:impl} we discuss the frozen-phonon implementation of phonon frequencies and EPI matrix elements in the context of the crystalline Gaussian basis of the PySCF package. In Section~\ref{sec:diamond} we show some preliminary results for the zero-point renormalization of diamond. We conclude this section with a summary of the challenges and our plans for addressing them.

\subsection{\textit{ab initio} electron phonon coupling}\label{sec:ai_epc}
Nearly all {\it ab initio} calculations of EPI use linear coupling:
\begin{equation}\label{eqn:EPC}
    \sum_{\mathbf{k}\mathbf{q}mnx} g_{(\mathbf{k} + \mathbf{q})n,\mathbf{k}m}^{\mathbf{q}x} c^{\dagger}_{(\mathbf{k} + \mathbf{q})n}c_{\mathbf{k}m} \left(b_{\mathbf{q}x} + b^{\dagger}_{-\mathbf{q}x}\right).
\end{equation}
Here, $m$ and $n$ label the electronic bands and $x$ labels the phonon branch. The EPI matrix elements are, in practice, computed as
\begin{equation}
	g^{x}_{pq} = \sum_{\alpha,s} \sqrt{\frac{\hbar}{2m_{s}\omega_{x}}} \epsilon^{x}_{s\alpha} \Big\langle p \Big|\frac{dV_{KS}}{dR_{s\alpha}} \Big|q \Big\rangle
\end{equation}
where we have suppressed the momentum indices in this expression. Here, $V_{KS}$ is the Kohn-Sham (or Hartree-Fock) potential, $s$ labels a particular atom, $\alpha$ labels a Cartesian direction, $m_s$ is the mass of the $s$th atom, $\omega_x$ is the frequency of the $x$th phonon mode, and the $\epsilon$ tensor transforms between Cartesian displacements and displacements in the phonon basis. In using a Hamiltonian of this form, there are two approximations. The first is that higher order terms, like the term quadratic in the displacements, are ignored. This approximation can be relaxed in principle by including higher order couplings. The second, less obvious approximation is due to the fact that the phonon frequencies come from a calculation which already includes, to some extent, the response of the ground state electronic energy to changes in the nuclear positions. This issue is discussed in more detail in Ref.~\onlinecite{VanLeeuwen2004}. Relaxing this approximation is difficult. One option would be to work within the self-consistent field-theoretic framework of the Hedin-Baym equations\cite{Baym1961,Hedin1970}. As a starting point, we use the standard linear coupling. 

\subsection{Implementation}\label{sec:impl}

In order to test this coupled cluster method in an {\it ab initio} setting, we have implemented the first-order electron phonon matrix for molecules and extended systems in the PySCF program package\cite{Sun2020}. The molecular implementation computes the analytical EPI matrix through the coupled-perturbed self-consistent field (CPSCF) formalism, similar to the implementation in FHI-AIMS\cite{Shang2017}. The periodic system implementation is based on a finite difference approach and currently supports only a single k-point. Specifically, finite differentiation is first performed on analytical nuclear gradients to yield the mass weighted hessian (dynamical matrix). Phonon modes are then obtained by diagonalizing this matrix. 

Throughout this work, we have used GTH-Pade pseudopotentials\cite{Goedecker1996,Hartwigsen1998} and the corresponding GTH Gaussian bases.\cite{VandeVondele2007} All integrals are generated by Fast Fourier transform based density fitting (FFTDF)\cite{VandeVondele2005}. In Table~\ref{tab:benchmark2} we compare the optical phonon frequency computed at the $\Gamma$ point using different basis sets. Amid the discrepancies in basis sets, pseudopotentials, and other numerical cutoffs, our results show overall good agreement with the implementations in CP2K\cite{Kuhne2020} and the plane-wave (PW) code Quantum Espresso (QE)\cite{Giannozzi2009}.
\begin{table}[h]
    \centering
    \begin{tabular}{c|ccc}
    \hline\hline
        $\omega^{OP}_{\Gamma}$ & PySCF & CP2K & QE \\
        \hline
        GTH-SZV(LDA) &  2385.56 & 2393.30 & -\\
        GTH-DZVP(LDA) & 2207.67 & 2214.58 & -\\
        GTH-TZVP(LDA) & 2290.95 & 2197.48& 2262.67(PW)\\
        GTH-SZV(PBE) &  2379.07 & 2384.69 & -\\
        GTH-DZVP(PBE) & 2202.70 & 2209.07 & -\\
        GTH-TZVP(PBE) & 2288.15 & 2191.82 & 2255.60(PW)\\
        \hline\hline
    \end{tabular}
    \caption{A comparison of the $\Gamma$ point optical phonon mode $(cm^{-1})$ from our implementation in PySCF against those computed using CP2K and QE. For our TZVP calculations, basis Gaussians with exponents less than 0.1 are discarded due to the diffuse nature of the functions. This could account for small discrepancies with CP2K in this basis set. Note that PySCF and CP2K use the same GTH pseudopotentials, while Hartwigsen-Goedeker-Hutter (HGH) pseudopotentials\cite{Hartwigsen1998} were used for QE. The QE calculations use 
    a kinetic energy cutoff of 60 Rydberg. To ensure that the QE and PySCF numbers can be directly compared, the QE calculations used the electron density from a $\Gamma$ point DFT calculation (unconverged with respect to Brillouin zone sampling) in the subsequent DFPT computation.}
    \label{tab:benchmark2}
\end{table}
Experimentally, the optical phonons of diamond appear around 1300 cm$^{-1}$ and this is consistent with calculations in large supercells (see for example Refs.~\onlinecite{Watanabe2004,Ishioka2006}). This shows the significant finite-size error associated with the 1x1x1 cell, but does not affect the comparison between different codes.

The evaluation of the Kohn-Sham response matrix is broken into three terms:
\begin{widetext}
\begin{equation}
    	\Big\langle p \Big|\frac{dV_{KS}}{dR_{s\alpha}} \Big|q \Big\rangle = \frac{d}{dR_{s\alpha}}\Big\langle p \Big|V_{KS} \Big|q \Big\rangle - \Big\langle \frac{dp}{dR_{s\alpha}} \Big|V_{KS} \Big|q \Big\rangle - \Big\langle p \Big|V_{KS} \Big|\frac{dq}{dR_{s\alpha}} \Big\rangle .
\end{equation}
\end{widetext}
The first term is evaluated by finite difference. The second and third term are obtained analytically as part of the nuclear gradient routine. In our implementation, the response matrix is evaluated in the AO basis and then transformed to the MO basis when needed. This is to avoid problems arising from different MO gauges that can occur in finite-difference calculations. Our implementation differs from standard PW codes in that the electron density and MO basis are converged in the same SCF procedure. 

To allow for easier comparison of our implementation against PW based codes, we take the occupied block of the potential response matrix as 
\begin{equation}
    Z^{s\alpha}_{ij} = \Big\langle i \Big|\frac{dV_{KS}}{dR_{s\alpha}} \Big|j \Big\rangle
\end{equation}
and define a gauge and basis independent $z$ metric for comparisons:
\begin{equation}
    z=\Tr{Z^{\dagger}Z}
\end{equation}

In Table~\ref{tab:benchmark} we compare our results for the $z$ metric with those from a PW implementation. For the PW reference, DFT/DFPT results from QE are used by Perturbo\cite{Zhou2020} to extract the potential response matrix. For our Gaussian basis implementation, we observe a slow basis convergence behavior moving from DZVP to TZVP, but again, given the differences in many numerical choices, our results in the TZVP basis are qualitatively similar to those from the PW reference.

\begin{table}[h]
    \centering
    \begin{tabular}{c|cc}
    \hline\hline
        $z$  & LDA & PBE \\
        \hline
        GTH-SZV  & 0.0864  & 0.0841 \\
        GTH-DZVP & 0.1639 &  0.1631\\
        GTH-TZVP & 0.1768 & 0.1739\\
        PW & 0.2278 & 0.2260\\
        \hline\hline
    \end{tabular}
    \caption{$z$ metric ($E_{\mathrm{h}}$) of diamond computed in a cGTO basis (PySCF) compared
    to results from QE/Perturbo computed in a PW basis.}
    \label{tab:benchmark}
\end{table}

In order to enable large simulations using ab initio Hamiltonians, the following strategies are adopted to optimize our ep-CC Python implementation: 
\begin{enumerate}
    \item We take advantage of the Symtensor library\footnote{https://github.com/yangcal/symtensor} to obtain an implicitly unrestricted implementation starting from the generalized spin-orbital equations from our code generator. The automatic use of symmetry in Symtensor is described in Ref.~\onlinecite{Gao2020a}.
    \item The Cyclops Tensor Framework\cite{Solomonik2013} is used as the numerical backend for tensor contraction to enable parallel computation.
\end{enumerate}

\subsection{Results: Diamond}\label{sec:diamond}
Diamond has emerged as a paradigmatic example in the field of {\it ab initio} electron-phonon computation, and the accurate computation of relatively simple quantities, like the zero-point renormalization (ZPR) (the shift of the bandgap due to phonon effects) remains a challenge. Experimental values based on isotopic shifts suggest a ZPR of the indirect gap of -364 meV\cite{Cardona2005a}. Calculations of the ZPR of the direct gap suggest that it is higher, closer to -600 meV\cite{Ramirez2006,Giustino2010,Antonius2014}. Importantly, it has been shown that many-body electronic effects are important to the ZPR of the direct gap\cite{Antonius2014} and that dynamical effects are important to capture some qualitative features of the EPI\cite{Cannuccia2011}. Some theoretical and experimental results are shown in Table~\ref{tab:diamond_ref}.
\begin{table*}
    \centering
    \begin{tabular}{c|c|c|c|c|c} \hline\hline
        ZPR &  EPI & electronic structure& ZPR& gap & reference\\ \hline
        -700 & - & tight-binding & PIMC&direct  & \onlinecite{Ramirez2006} \\
        -615 & LDA & LDA& AHC& direct& \onlinecite{Giustino2010}\\
        -628 & LDA & GW & AHC& direct& \onlinecite{Antonius2014} \\
        -334 & - & LDA& Ref.~\onlinecite{Monserrat2014}& indirect& \onlinecite{Monserrat2014} \\
        -345 & - & LDA& WL& indirect& \onlinecite{Zacharias2016} \\
        -337 & - & GW & MC& indirect & \onlinecite{Karsai2018} \\
        -364 & - &Experiment& - & indirect & \onlinecite{Cardona2005a}\\ \hline\hline
    \end{tabular}
    \caption{Selected literature results for the ZPR of diamond. Monte Carlo is abbreviated as MC. Path integral molecular dynamics is abbreviated as PIMD, Allen-Heine-Cordona\cite{Allen1976,Allen1981} theory is abbreviated as AHC, and the theory of Williams\cite{Williams1951} and Lax\cite{Lax1952} is abbreviated as WL. The method used to get the ZPR in Ref.~\onlinecite{Monserrat2014} does not have a commonly used name, but it is clearly described in given reference.}
    \label{tab:diamond_ref}
\end{table*}

In Table~\ref{tab:diamond} we show the ZPR of diamond computed by IP/EA-EOM-ep-CCSD-1-S1 and IP/EA-EOM-CCSD-PT2. The EPI matrix elements and phonon frequencies are computed from mean-field Hartree-Fock calculations. It was necessary to remove the most diffuse s orbital from the GTH-DZVP basis and the most diffuse s and p orbitals from the GTH-TZVP basis in order to eliminate numerical instabilities in the calculation of the EPI matrix elements.
The experimental lattice constant of diamond, 3.566\AA, is used throughout. For EOM-CCSD-PT2, the electronic CCSD amplitudes are used along with a PT2 estimate of the electron-phonon amplitudes:
\begin{equation}
    t_{i,x}^a = -\frac{g_{i,x}^a}{\varepsilon_a - \varepsilon_i + \omega_x}.
\end{equation}

The quantities in Table~\ref{tab:diamond} are directly comparable to the ZPR of the direct gap which has recently been reported to be in the range of -600 to -700 meV\cite{Ramirez2006,Giustino2010,Antonius2014}. However, the very small size of our simulation cell means that these numbers require some estimate of the finite-size error to be meaningfully compared with experiment. In diamond, the finite size effects are significant. However, the strength of coupled cluster methods is that they explicitly treat many-body electronic effects as well as dynamical electron-phonon correlation in a consistent framework. Thus recomputation using the approximate literature treatments within the same smaller cells would allow for the magnitude of higher-order many-body effects to be estimated from these CC calculations. 
\begin{table}[h]
    \centering
    \begin{tabular}{c|cccc}
    \hline\hline
        Basis & \multicolumn{2}{c}{CCSD-1-S1} & \multicolumn{2}{c}{CCSD-PT2} \\
         & full & no-VV &  full & no-VV \\
        \hline
        GTH-SZV &  -671& -366& -671 & -366\\
        GTH-DZVP$^{\ast}$ & -831 & -617& -826& -512\\
        GTH-TZVP$^{\ast}$ & -1343 & -767& -1115& -645\\
        \hline\hline
    \end{tabular}
    \caption{Band gap renormalization (meV) at the $\Gamma$ point (direct gap) for a 1x1x1 unit cell in different basis sets. Note that the most diffuse s-orbital was removed from the GTH-DZVP basis and the most diffuse s and p orbitals were removed from the GTH-TZVP basis. In the ``no-VV" columns, the unoccupied-unoccupied EPI matrix elements were ignored. This provides a more direct comparison with typical treatments of band-gap renormalization.}
    \label{tab:diamond}
\end{table}

We can draw two conclusions from these finite-size ep-CC calculations. First, using the PT2 estimate of the coupled amplitudes provides EOM results that are similar to the converged CC results, but the converged CC amplitudes provide EOM results that are systematically lower. This suggests that the converged CC ground state is probably not necessary to obtain reasonable excited-state properties of typical large-gap insulators. Second, we find that the band-gap renormalization becomes unexpectedly large as the size of the basis set is increased. This affect appears to be largely due to the unoccupied-unoccupied (virtual-virtual, or VV) block of the electron-phonon matrix elements which do not appear in the widely used Allen-Heine-Cordona (AHC) treatment\cite{Allen1976,Allen1981}. This could indicate that the Hamiltonian of Equation~\ref{eqn:EPC} does not properly describe the electron-phonon coupling between unoccupied bands which does not enter into typical calculations. Alternatively, it could be due to the small finite size of the simulation.

Results for a larger supercell are shown in Table~\ref{tab:diamond2}. 
\begin{table}[]
    \centering
    \begin{tabular}{c|cc}
    \hline\hline
         supercell & CCSD-1-S1 & CCSD-PT2\\ \hline
         1x1x1 & -671& -671\\
         2x2x2 & -134& -142\\
         3x3x3 & - & -42\\ 
         \hline\hline
    \end{tabular}
    \caption{ZPR (meV) of diamond supercells in the GTH-SZV basis set. The 2x2x2 and 3x3x3 supercells provide estimates of the indirect band gap renormalization. In the 3x3x3 supercell we were unable to obtain converged CCSD-1-S1 amplitudes.}
    \label{tab:diamond2}
\end{table}
These results are not constrained to compute the direct gap, so the results for 2x2x2 and 3x3x3 supercells should be viewed as finite-size approximations to the ZPR of the indirect bandgap. These results affirm that using CCSD-PT2 amplitudes in the EOM calculation is a reasonable approximation. The ZPR is smaller for larger supercells which is consistent with the smaller ZPR for the indirect gap. Though the simulation cell is still too small for a reliable extrapolation, the numbers are consistent in magnitude with results that have been reported in the literature. The slow and oscillatory convergence of the ZPR of diamond with supercell size is a well-known problem\cite{Monserrat2014,Ponce2015,Zacharias2016}.

\subsection{Future directions for {\it ab initio} calculations}
In the previous section, we identified two significant sources of error in our CC calculations which explicitly include EPI. The first is the finite-size error which is difficult to control when the CC equations must be solved simultaneously for all electronic and phononic degrees of freedom. The second is the form of the EPI term itself which may be insufficient, especially for the unoccupied bands.

We intend to address the finite-size error by using a perturbative correction to EOM-CCSD eigenvalues which can be interpolated to denser k-point grids as is usually done in traditional calculations of EPI. The coupled cluster framework presented here will be useful in evaluating the validity of these perturbative approximations.

The validity of the linear EPI term also needs to be investigated further. This requires very accurate calculations on small systems or model systems, and we expect this coupled cluster framework to be useful in that it can provide more systematic results for such problems.

\section{Conclusions}
We have presented a coupled cluster framework for a systematic, correlated treatment of interacting electrons and phonons. The theory is a straightforward combination of fermionic (electronic) and bosonic (phononic) coupled cluster ansatze. Despite the formal simplicity of the ansatz, sophisticated diagrammatic techniques and automated operator algebra were necessary to efficiently implement the equations. These techniques are described in the appendices. In order to benchmark these methods, we have applied them to the Hubbard-Holstein model. Calculations on the 4-site HH model, which can be exactly solved numerically, reveal that all the CC methods discussed here perform well for small to moderate coupling. Calculations of the excited-state properties of the model suggest that the EOM-ep-CC methods can provide excited state energies with an accuracy comparable to EOM-CC for electronic excitations.

Finally we have discussed the details of an {\it ab initio} implementation in the context of crystalline Gaussian-type orbitals. Preliminary calculations on the ZPR of diamond are consistent with values reported in the literature, but a better treatment of finite-size error is necessary for truly quantitative calculations. This motivates the future development of more approximate theories that can utilize EPI matrix elements interpolated onto a very fine momentum-space grid. We found unexpectedly large values for the ZPR when coupling between virtual bands was included in the calculation which suggests that the approximate, linear form of the EPI may not be sufficient in the more sophisticated many-body treatments of electron-phonon effects where these states must enter.

\begin{acknowledgments}
We thank Jinjian Zhou for helpful discussion in implementation of {\it ab initio} EPI matrix using cGTO basis. A. F. W. and G. K. C. acknowledge support from the US Department of Energy via the M2QM EFRC under award no. de-sc0019330. Y. G. and A. J. M. acknowledge the support of ONR under Grant No. N00014-18-1-2101.
\end{acknowledgments}

\appendix

\section{Diagrammatic derivation of ep-CCSD-1-S1 equations}\label{app:diagram}
\begin{figure}
    \centering
    \includegraphics[scale=0.5]{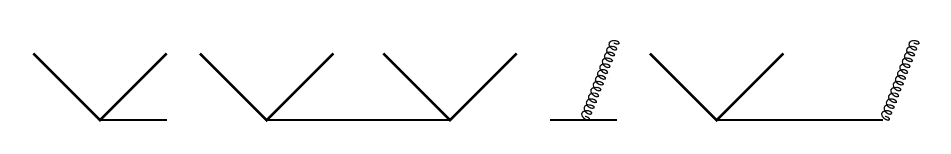}
    \caption{Diagrammatic representation of the electronic $T_1$, electronic $T_2$, phononic $T_1$, and the lowest order piece of $T_{\mathrm{ep}}$ respectively. These are the amplitudes of the ep-CCSD-1-S1 method. The arrows indicating the direction of the fermion propagators have been omitted in this case since they can be inferred from the skeletons.}
    \label{fig:ep_cc_amp}
\end{figure}

The equations for ep-CC methods can be derived using a diagrammatic language just as for traditional fermionic CC. As usual, we use solid arrows for fermion propagators, but we must also consider phonon propagators which we indicate with a spring-like line (often used as a gluon line in high energy physics). For example, the diagrammatic representations of the ep-CCSD-1-S1 amplitudes are shown in Figure~\ref{fig:ep_cc_amp}.

\begin{figure}
    \centering
    \includegraphics[scale=0.45]{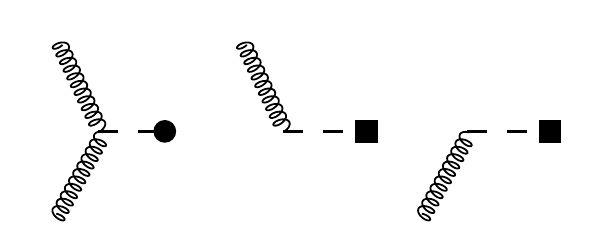}
    \includegraphics[scale=0.45]{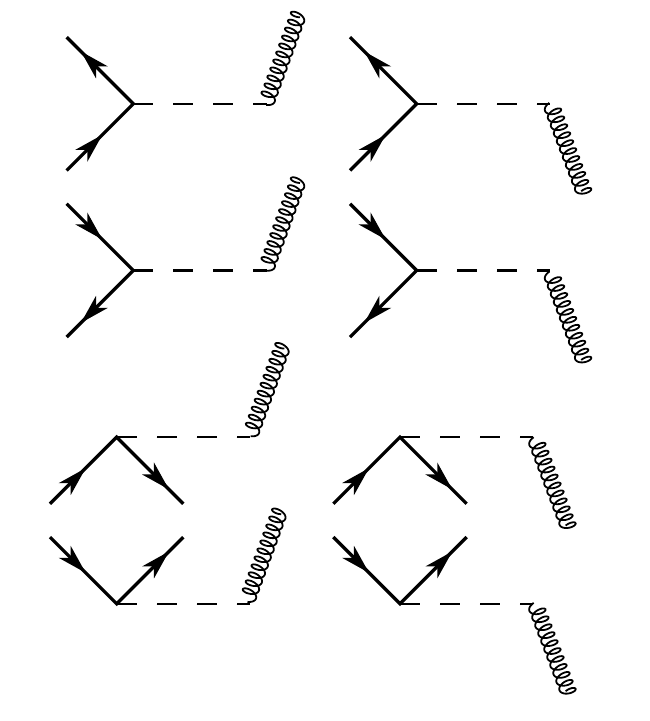}
    \caption{Diagrammatic representation of the phonon and electron-phonon Hamiltonian respectively. The diagonal harmonic part of of the Hamiltonian ($\omega$) is represented by a circle and the single phonon part ($G$) is represented by a square.}
    \label{fig:Hep}
\end{figure}

\begin{figure}
    \centering
    \includegraphics[scale=0.5]{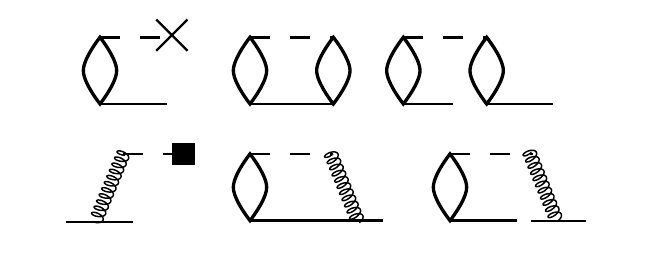}
    \caption{Diagrammatic contributions to the ep-CCSD-1-S1 energy. The first three terms (top row) are the energy diagrams familiar from fermionic coupled cluster.}
    \label{fig:energy}
\end{figure}

We assume a normal-ordered Hamiltonian of the form
\begin{align}
    H &= \sum_{pq}f_{pq}N[a_p^{\dagger}a_q] \nonumber \\ 
    &+ \frac{1}{4}
    \sum_{pqrs}\langle pq || rs\rangle N[a_p^{\dagger}a_q^{\dagger}a_sa_r] \nonumber \\
    &+ \sum_x \omega_x N[b_x^{\dagger}b_x] \nonumber \\
    &+ \sum_x G_x\left(N[b_x] + N[b_x^{\dagger}]\right)
    \nonumber\\
    &+ \sum_{pq,x}g_{pq}^xN[a_p^{\dagger}a_q(b_x + b_x^{\dagger})],
\end{align}
where we have used $N[\ldots]$ to indicate normal ordering.
Even if a linear pure-phonon term, $G$, does not appear in the Hamiltonian it can arise from normal-ordering of the electron-phonon term with respect to the Fermi vacuum. In addition to the Fock matrix and electron-electron interaction we use the diagrammatic components shown in Figure~\ref{fig:Hep} to represent these additional terms in the Hamiltonian.
Given these components, and armed with the connected cluster property of the similarity transformed Hamiltonian, the energy and amplitude equations can be derived in the same manner as in fermionic CC (see Chapters 4,5,9, and 10 of Ref.~\onlinecite{Shavitt2009}). As an example, we show the diagrammatic contributions to the energy in Figure~\ref{fig:energy}.

\section{Automatic code generation for ep-CC equations}
\label{app:codegen}
In order to verify our diagrammatic derivations for ep-CCSD-1-S1 and to easily implement more complicated theories, we have used a code generator which we have made available on github.\footnote{https://github.com/awhite862/wick} Our code generator provides a Python interface and evaluates expressions of the form
\begin{equation}
    \langle \Phi_0, 0| [\ldots]|\Phi_0, 0\rangle ,
\end{equation}
where $\Phi_0$ is a Fermi vacuum, $0$ is the Boson vacuum, and $[\ldots]$ can be expressions constructed from second-quantized fermion and boson operators. The evaluation is accomplished by algebraic application of Wick's theorem followed by simplification.

This algebraic approach has been used for fermionic\cite{Janssen1991,Hirata2003} and bosonic theories\cite{Faucheaux2015}, but we are not aware of analogous software for the coupled problem. This approach allows the code generator to be as general as possible and provides a validation that is completely independent of the diagrammatic approach described in Appendix~\ref{app:diagram}. The equations are available on github\footnote{https://github.com/awhite862/gen\_epcc}.

\section{Equations for the EOM sigma vector}\label{app:EOM}
Our code generator was also used to derive equations for the EOM sigma vector. The naive approach is to directly evaluate an expression of the form
\begin{equation}
    \sigma_{i,x}^a = \langle \Phi_0, 0| a_i^{\dagger}a_ab_xe^{-T}
    He^TR|\Phi_0, 0\rangle .
\end{equation}
While this approach provides the correct equations, the resulting equations contain hundreds of terms, even for simple EOM theories, and are therefore difficult to optimize.

The approach we have taken is to separately generate the different sectors of the similarity transformed Hamiltonian, $\bar{H}$. We first generate the equations for
\begin{equation}
    \bar{H}_{\mu\nu} = \langle \Phi_0, 0| \mu \bar{H}\nu^{\dagger}|\Phi_0, 0\rangle
\end{equation}
for all relevant $\mu$ and $\nu$. Next we generate equations for the different sectors of the configuration-interaction-like sigma vector as
\begin{equation}
    \sigma_{\mu} = \sum_{nu} H_{\mu\nu}R_{\nu}.
\end{equation}
Finally, these equations are manually optimized for memory usage and computation time. This process of generating separately the equations for the relevant blocks of $\bar{H}$ closely mirrors optimal implementations of EOM which greatly simplifies the optimization process. The unoptimized equations are avaible on github.\cite{Note3}.

\section*{References}
\bibliography{references}

\end{document}